\documentclass{article}
\usepackage{amsfonts}

\usepackage{eurosym}
\usepackage{amsmath,amssymb,epsf,epsfig,amsthm,bm,graphicx,subfigure}

\begin{document}
\title{Uses of Chern-Simons actions \footnote{Talk given at the conference \textit{Ten Years of AdS/CFT}, a workshop to celebrate the tenth anniversary of the Maldacena Conjecture. Buenos Aires, December 2007. }}
\author{Jorge Zanelli \\
\large{Centro de Estudios Cient\'{\i}ficos (CECS), Valdivia, Chile}}
\maketitle
\begin{abstract}
The role of Chern-Simons (\textbf{CS}) actions is reviewed, starting from the observation that all classical actions in Hamiltonian form can be viewed
as 0+1 CS systems, in the same class with the coupling between the electromagnetic field and a point charge. Some suggestions are derived that
could shed some light on the quantization problem in $D=2n-1>3$ dimensions.
\end{abstract}

\section{Introduction}
Chern Simons actions describe the boundary dynamics of the simplest gauge theories: topological field theories defined by characteristic classes, like
the Pontryagin and the Euler forms. These intriguing field theories exist in even dimensional spacetimes and have no local propagating degrees of freedom.
Their actions are stationary under arbitrary continuous variations of the field (a gauge connection in some Lie algebra), provided their values are
kept fixed at the boundary. In this sense, the relation between a topological field theory and its corresponding CS dynamics at the boundary, could be viewed as an example of the AdS/CFT correspondence we are celebrating at this conference.

\subsection{A bit of history}
Chern-Simons forms provide Lagrangians for gauge theories, invariant under some symmetry group ($G$) in a certain odd-dimensional manifold $M$. The main difference between CS and more conventional Yang-Mills lagrangians is that they are written explicitly as functions of a connection $\mathbf{A}$ and
its exterior derivatives, but they \textit{cannot} be written as local functions involving only the curvature, $\mathbf{F}=d\mathbf{A}+\mathbf{A}\wedge\mathbf{A}$.

In physics, CS forms were originally encountered in the discussion of chiral anomalies in four spacetime dimensions, which signal the violation of the
classically conserved chiral currents by quantum mechanical effects. By direct computation, the deviation from the classical conservation law (anomaly) was shown to be proportional to $C_4(\mathbf{F})$, the Chern class for the connection that couples to the chiral current \cite{Schonfeld,Deser-Jackiw-Templeton}. It was later observed that this form could be written as the exterior derivative of a local three form, $C_4(\mathbf{F})=d\mathcal{C}_3(\mathbf{A})$, where $\mathcal{C}_3(\mathbf{A})$ is a function of the connection, originaly discussed by Chern and Simons in the mathematical literature \cite{Chern+Simons}. For a historical overview, see \cite{Deser98}.

It seems that a CS form was used as a Lagrangian for the first time in the 11-dimensional supergravity model of Cremmer, Julia and Scherk, where the
action contains a CS term for a three-form field needed by supersymmetry \cite{Cremmer:1978km}. It was later realized that CS forms define reasonable
--and potentially useful-- Lagrangians for field theories in three spacetime dimensions. CS actions have also been invoked for the description of the
quantum Hall effect \cite{Balachandran:1995dv} and are related to the polynomial invariants of knot theory \cite{Witten:1988hf}; CS theory is
also closely related to conformal field theory and the Wess-Zumino-Witten (WZW) action in two dimensions \cite{Witten:1988hf,CS+}. The CS action has
also been used for describing the gluon plasma in QCD \cite{Efraty:1992gk}. Finally, the standard gravity theory in 2+1 dimensions itself was shown to
be a CS system \cite{Achucarro:1987vz,Achucarro:1989gm,Witten:1988hc}. Moreover, CS forms in more dimensions can describe gravities or
supergravites which are genuine gauge theories \cite{Chamseddine:1990gk} (for a review, see \cite{Zanelli:2005sa}).

Examples of CS actions, however, have existed in the physical literature much longer that this. In fact, the electromagnetic coupling between a point
charge in an external electromagnetic field, $\dot{z}^{\mu}A_{\mu}$, defines a one-dimensional CS system. Remarkably, any mechanical system with a finite number of degrees of freedom written in Hamiltonian form is also a $0+1$ CS system \cite{Dunne-Jackiw-Trugenberger,Saavedra}. It is therefore fair to say that all of classical mechanics can be viewed as the study of a class of CS systems. The extension of this assertion to infinite-dimensional hamiltonian systems, however, is not straightforward as the passage from mechanics to field theory is highly nontrivial. 

\subsection{0+1-dimensional CS theory}
The coupling of an electrically charged point particle with an external electromagnetic field, 
\begin{equation}  \label{eA}
I= e \int_{\Gamma} A_{\mu}(z)dz^{\mu},
\end{equation}
has some peculiar features: it does not make reference to the metric of the spacetime where the interaction takes place; although it depends explicitly
on a non-invariant gauge field, it is gauge invariant, provided charge conservation is verified in the system. This unique form of interaction term is an expression of the minimal coupling between charged matter and the electromagnetic field, implemented through the substitution of ordinary derivatives by (gauge) covariant derivatives. This form of interaction in turn means that the field that mediates the electromagnetic interaction is a connection on a fiber bundle (gauge theory). This form is common to all four basic interactions in nature, and as we shall see below, is also the simplest example of a CS system.

The action (\ref{eA}) can be viewed as a functional of $A$ or as a functional of the embedding coordinates $z^{\mu }$. In the first case, it might seem to lead to an inconsistent classical system: the Euler-Lagrange equation obtained varying with respect to $A$ reads $1=0$. However, this is a very narrow interpretation of the variation. In fact, what one obtains is  
\begin{equation}
\delta I[A]=e\int_{\Gamma }\delta A_{\mu }(z)dz^{\mu }=0,
\end{equation}
which only states that the integrand must be an exact form, 
\begin{equation}
\delta A(z)=d\alpha (z),  \label{deltaA}
\end{equation}
with $\alpha (z_{+})=\alpha (z_{-})$, where $\{z_{+},z_{-}\}=\partial \Gamma$. Thus, the only local requirement on the field $A$ derived from the
stationarity of the classical action is that it can only vary by an exact form, but is otherwise arbitrary. In other words, the classical equation only informs us of the fact that $A$ is an abelian connection, which need not be an obvious fact from reading (\ref{eA}). On the other hand, the periodicity of $\alpha (z)$ can be automatically guaranteed if the manifold $\Gamma $ is closed on itself, i. e., if it has no boundary, $\partial \Gamma=0$. So, the action principle hints to the fact that the coordinate $z$ along $\Gamma $ should be periodic.

\subsubsection{Classical mechanics as a 0+1 CS system}
There is an interesting interpretation of equation (\ref{eA}) if considered as a functional of $z^{\mu}$, the embedding coordinates for $\Gamma$ in a
higher-dimensional target manifold. This is the conventional interpretation leading to equations of motion for a charge in the presence of an external
e-m potential $A$, although in this case we have not included a kinetic term for the charge.

In the hamiltonian form, the action for a mechanical system of a finite number of degrees of freedom reads 
\begin{equation}
I[p,q]=\int_{\Gamma }[p_{i}dq^{i}-H(p,q)dt],\;\;\;i=1,2,3,...,s,
\label{Ham-action}
\end{equation}
and can be seen to be of the form (\ref{eA}) if one identifies $A_{i}=p_{i}$, $A_{s+i}=0$, $A_{0}=-H$, and $z^{\mu }=(t,q^{i},p_{j})$. In other
words, the hamiltonian action is equivalent to the 0+1 CS form in a $(2s+1)$-dimensional embedding space $M^{2s+1}$. The analogy can be made more
apparent if the kinetic term is written in the skew-symmetric form, $(p_{i}dq^{i}-q^{i}dp_{i})/2$ (the corresponding identification is now $A_{i}=p_{i}/2$, $A_{s+i}=-q^{i}/2$). Note that the phase space, enlarged by the inclusion of time, is odd-dimensional, and Hamilton's equations can be written as 
\begin{equation} 
F_{ab}\dot{z}^{b}=E_{a},\;\;\;a,b=1,2,3,...2s,  \label{Ham-eqs}
\end{equation}
where we have defined $E_{a}=F_{0a}=\partial_{0}A_{a}-\partial_{a}A_{0}=\partial _{a}H$, and the field strength $F_{ab}$ is identified as the symplectic form. Thus, we conclude that all of classical mechanics can be understood as a CS system where Hamilton's equations describe a particle moving under the influence of an external electromagnetic field that produces zero net Lorentz force on the charge ($e[\overrightarrow{E}+ \overrightarrow{v} \times \overrightarrow{B}]=0$). One application of the analogy between the dynamics of classical mechanics and electromagnetic phenomena is Feynman's derivation of Maxwell's equations from classical mechanics \cite{Dyson}.

But, where is gauge invariance in a mechanical system? A moment's thought reveals that, in this language, gauge invariance is just the invariance of
the Euler-Lagrange equations under the addition of a total derivative to the lagrangian, $L\rightarrow L+d\Omega(p,q,t)$, where $L=p_idq^i - H(p,q)dt$.
In other words, gauge invariance in this framework is the symmetry of classical mechanics under canonical transformations.

In standard (unconstrained) hamiltonian systems, $F_{ab}$ is a nondegenerate two-form and its inverse $F^{ab}$ is defined everywhere in phase space. If $F_{ab}$ has constant rank $2r<2s$, its inverse exists on a submanifold of dimension $2r$, as it occurs in constrained systems that possess some form of gauge invariance. If the rank is not constant, the system belongs to a degenerate class that possess different number of degrees of freedom throughout phase space \cite{Saavedra}. Higher dimensional ($D\geq5$) CS systems generically belong to this latter class of degenerate systems.

\subsubsection{Boundary conditions}
Let's rewrite the action (\ref{eA}) identifying the time-like coordinate $z^0$ of $M^{2s+1}$ with the time parameter of the curve $\Gamma$. 
\begin{equation}  \label{InR}
I[z]= e \int_{\Gamma} [A_0(z) + A_a(z)\dot{z}^a]dt.
\end{equation}

As usual, the equation of motion (\ref{Ham-eqs}) was obtained extremizing the action under arbitrary variations $\delta z(t)$ for $t_1 < t < t_2$, and dropping a boundary term, viz., 
\begin{equation}  \label{BT}
\delta I[z]= \int_{t_1}^{t_2} \delta z^a[F_{ab}\dot{z}^b-E_a]dt+ \left.
\delta z^a A_a(z) \right|_{t_1}^{t_2}.
\end{equation}
In order for this extremum to be well defined, the second term in the r.h.s. of (\ref{BT}) must also vanish. The question is, under what boundary conditions is the variation performed? Namely, what conditions must be imposed on $z$ at $t_1$ and $t_2$ so that this last term vanishes?

One might be inclined to impose Dirichlet boundary conditions: fixing the values of $z$ at both ends, $z(t_1)=z_1$ and $z(t_2)=z_2$. However, there is something excessive about this prescription because (\ref{Ham-eqs}) is a first order equation for $z$, and therefore one cannot in general find a solution for fixed values of $z$ at both end points. In fact, the entire solution $z_a(t)$ can be found given the values of the coordinates $z^a$ at one instant $t_0$. Another form of the same problem is that (\ref{InR}) is actually a phase space representation where the $2s$ coordinates are twice as many as degrees of freedom in the system (assuming no additional first class constraints), and by definition, this is also the number of initial conditions needed to specify the solution uniquely. In quantum mechanics, the wave function cannot be defined specifying all the phase space coordinates of a particle at both ends of a trajectory, as that would violate the uncertainty principle. Thus, an action principle in which the phase space coordinates are fixed at both end points is incompatible both with the data required for the integration of the classical equations of motion, and with quantum mechanics.

Actually, the minimal requirement for the vanishing of the second term in the r.h.s. of (\ref{BT}) is a weaker condition on $\delta z^a$ and $A(z)$, 
\begin{equation}  \label{bc}
\delta z^a(t_1) A_a(z(t_1)) = \delta z^a(t_2) A_a(z(t_2)).
\end{equation}
This looks like an extreme case of fine tuning between the values of $z^a$ and their variations at two points in the trajectory. One way of achieving
this without fine tuning is by assuming the trajectory to be periodic, $z^a(t_1)=z^a(t_2)$, and therefore 
\begin{equation}  \label{PBC}
\delta z^a(t_1)= \delta z^a(t_2), \;\;\; A(z(t_1))= A(z(t_2)).
\end{equation}
An equivalent result is obtained for antiperiodic boundary conditions, $z^a(t_1)= -z^a(t_2)$, provided $A(-z)=- A(z)$, a possibility that occurs naturally in fermionic systems \cite{Galvao:1980cu}.

\subsection{Constrained dynamics}
In order to analyze the quantum behavior of CS systems, one would like to apply Dirac's hamiltonian analysis \cite{DiracLectures} to (\ref{InR}). From
the definition of momenta, a set of primary constraints is found, 
\begin{equation}  \label{primary}
\Phi_a\equiv p_a-A_a(z) \approx 0\;, a=1, 2, ...,2s.
\end{equation}
The canonical hamiltonian is $H_c=-A_0(z)$, and the extended hamiltonian takes the form 
\begin{equation}  \label{extHam}
H= -A_0(z) + \lambda^a\Phi_a(z,p).
\end{equation}
Preservation of the constraints, $\Phi_a=0$ under the time evolution generated by $H$ requires 
\begin{equation}  \label{phi-dot}
\dot{\Phi}_a=F_{ab}\lambda^b - E_a \approx 0 \; ,
\end{equation}
where $F_{ab}=\partial_a A_b - \partial_b A_a$, and $E_a = -\partial_a A_0$. This equation shows that the Lagrange multipliers $\lambda^a$ are
essentially the same as the velocity. The nature of these constraints (first/second class) depends on the rank $r$ of the matrix 
\begin{equation}  \label{F}
[\Phi_a,\Phi_b] = F_{ab}.
\end{equation}
This matrix is the symplectic form in Hamilton's equations (\ref{Ham-eqs}), and also the Dirac matrix of the second class constraints $\Phi_a \approx 0$. If this matrix has maximal rank, $r=2s$, then all $\Phi$'s are second class and by Darboux theorem there exist canonical coordinates for which $F_{ab}$ takes the canonical form, 
\begin{equation}  \label{Darboux}
F_{ab}= \left[ 
\begin{array}{cccccc}
0 & 1 &  &  &  &  \\ 
-1 & 0 &  &  &  &  \\ 
&  & 0 & 1 &  &  \\ 
&  & -1 & 0 &  &  \\ 
&  &  &  & \cdot &  \\ 
&  &  &  &  & \cdot \\ 
&  &  &  &  & 
\end{array}
\right].
\end{equation}

The fact that the Dirac matrix is invertible means that the second class constraints can be locally solved and half of the coordinates $z^a$ are momenta conjugate to the other half, so that the action (\ref{InR}) actually describes a system of $s$ degrees of freedom. 

It is often the case that the two-form $F$ has rank $r=2m<2s$. Then, some of the phase space coordinates are not dynamically determined and therefore have an arbitrary time evolution. In this case, the system possesses some form of gauge invariance and there is a choice of (canonical) coordinates in which $F$ adopts the form (\ref{Darboux}) in the $2m \times 2m$ upper diagonal block, with zeros elsewhere. The physical phase space corresponds then to a $2m$-dimensional submanifold of dynamically determined coordinates.

A more pathological situation occurs when the rank of $F$ is not constant throughout phase space. Then $\det F$ is a function of the phase space coordinates that vanishes on a set of points, $\Sigma$. Then it is impossible to write down $F_{ab}$ in canonical form in the neighborhood of $\Sigma$. This is the case of the degenerate dynamical systems, in which the system can evolve from an initial state with $m_1$ degrees of freedom to another with $m_2<m_1$ degrees o freedom in a finite time \cite{Saavedra}. A similar but unrelated pathology can occur in an action of the form (\ref{eA}) when the constraints are not regular throughout phase space \cite{Henneaux-Teitelboim}. In such cases, the constraint surface defined by the condition $\det F=0$ fails to be an everywhere differentiable manifold, as for instance when the surface intersects itself, or when it is composed of several intersecting differentiable surfaces \cite{Miskovic:2003tn,Miskovic:2003ex}.

\section{Quantum Mechanics}
Let us ignore for the time being the pathologies of degenerate and irregular systems, and formally apply the standard quantization approach of Feynman's sum over paths.

\subsection{Holonomy/flux quantization}
The path integral for the action (\ref{eA}) reads 
\begin{equation}  \label{PI}
Z=\int [dz] \exp \left[ \frac{i}{\hbar}e \int_{z_1}^{z_2} A_{\mu}dz^{\mu}\right],
\end{equation}
where the sum over paths includes all continuous trajectories between $z_1$ and $z_2$, and the measure of integration in path space includes the symplectic form, 
\begin{equation}
[dz]=\Pi_{\{i,t \}}\left( \det [F(z)] dz^i(t)\right).
\end{equation} 

As already mentioned, the trajectories for which the action principle is naturally adapted are closed (periodic) paths for which $z_1$ and $z_1$ are identified. Hence, the largest contributions to the path integral come from periodic orbits that have maximum constructive interference. Those are orbits $\Gamma$ for which the holonomies $\oint_{\Gamma} A$ are integer multiples of $2\pi$,
\begin{equation}  \label{BS}
\frac{1}{\hbar} e \oint_{\Gamma} A_{\mu}dz^{\mu}= 2n\pi.
\end{equation}
Since the paths $\Gamma$ are closed, each of them can be viewed as the boundary of a two-dimensional submanifold, $\Sigma$, and therefore (\ref{BS}) means that the fluxes of the field strength, are quantized, 
\begin{equation}  \label{Flux}
e \oint_{\Gamma} A=e \int _{\Sigma}F=2n\pi\hbar.
\end{equation}
Although there are infinitely many oriented, two-dimensional surfaces  $\Sigma$ for every $\Gamma$, it is reassuring to see that the holonomy quantization condition also implies that the flux of $F$ can be evaluated on any $\Sigma$, because the difference in flux of an exact two-form through two co-bordant surfaces is also quantized. This guarantees that the observed (i. e., gauge invariant) configurations are those with quantized holonomies as well.

\subsubsection{Bohr-Sommerfeld quantization}
When the previous result is applied to the classical mechanical case, it translates into 
\begin{equation}  \label{BS+}
\oint_{\Gamma} \left( p_i dq^i - Hdt \right) = 2n\pi\hbar,
\end{equation}
which are the Bohr-Sommerfeld quantization rules of the old quantum mechanics theory \cite{Messiah}. In this case, all the coordinates $z^a=\{q^i, p_i\}$ must be periodic in the time parameter $t=z^0$. But it is possible to consider the amplitude for the system going from a certain initial configuration at $t=0$ to some final state at $t=1$. Then, the paths that dominate the sum over histories are those which have maximal constructive interference. That means, the paths $\Gamma_k$ that contribute most to the amplitude along $\Gamma_0$, are those for which 
\begin{equation}  \label{Constr-Interf}
\int_{\Gamma_0} \left( p_i dq^i - Hdt \right) - \int_{\Gamma_k} \left( p_i dq^i - Hdt\right)= 2m(k)\pi\hbar,
\end{equation}
where $m(k)$ is some integer. The left hand side is an integral around the closed loop composed by $\Gamma_0$ and $\Gamma_k$. The reader can recognize
again in this expression the Bohr-Sommerfeld quantization condition.

\subsubsection{Electron-Monopole charge quantization}
The flux quantization presented above is also related to the monopole-electric charge quantization condition discovered by Dirac \cite{Dirac}. As already mentioned, the 0+1 CS action (\ref{eA}) can be used to describe both a mechanical system of finitely many degrees of freedom, or a particle moving in the presence of an external electromagnetic field. In the latter case, if the field $A$ corresponds to a point magnetic source of strength $g$, 
\begin{equation}  \label{Monopole}
A=\left\{%
\begin{array}{cc}
g(1-cos \theta)d\phi & 0\leq \theta \leq \pi/2 \\ 
-g(1+cos \theta)d\phi & \pi/2\leq \theta \leq \pi
\end{array}
\right.
\end{equation}
This 1-form describes a mechanical system in a two-dimensional phase space, composed by a particle of unit charge in the presence of a magnetic monopole
of strength $g$. The magnetic flux across a closed 2-sphere is 
\begin{eqnarray}
e \int_{S^2} F &=& eg \int_{S^2} \sin\theta d\theta \wedge d\phi \\
&=& 4\pi e g.  \label{S2flux}
\end{eqnarray}
By the previous discussion, this integral must be quantized and therefore, 
\begin{equation}  \label{quantumg}
eg=\frac{n}{2}\hbar\; ,
\end{equation}
which is yet another way to obtain Dirac's famous quantization condition.

\subsection{Lessons from 0+1 CS systems}
We have seen that all mechanical systems with a finite number of degrees of freedom are CS systems. The quantization rules appear as quantization of the flux of the symplectic form enclosed by the orbits in phase space. Dirac's quantization of the product of magnetic and electric charge is a particular case of the same quantization rule. In fact, that same condition arises in the quantization of the only free parameter of Chern-Simons theories in $2n+1$ dimensions, and in particular, in the quantization of the analogue of Newton's constant in CS gravities \cite{Zanelli:1994ti}.

In general, the quantization rules are the condition for maximal constructive interference (\emph{i. e.}, observability) of the holonomies in the extended $2s+1$ dimensional phase space plus time. Taking these ideas to higher dimensional objects (strings, membranes, $p$-branes), requires a classification of the holonomies produced by these higher dimensional objects in their evolution in phase space.

One can view the 0+1 CS action as the interacting part of the action for a charged particle in the presence of an external electromagnetic field. What is, from this point of view, the role of the kinetic term of the particle? 

The addition of a kinetic term, 
\begin{equation}  \label{k}
I[z]= \int_{\Gamma}\left[\frac{1}{2} \gamma_{ab}(z)\dot{z}^a \dot{z}^b \right]dt,
\end{equation}
to the full action modifies the path integral by assigning a different weight to the trajectories depending on their velocity. This can be viewed as a modification of the measure that regularizes the path integral by providing an exponential damping, or a cut off, at high velocities. If there is a static classical solution, it can be used as the ground state for the construction of a quantum theory. In that case, the small fluctuations have a Gaussian spectrum characterized by the kinetic term and the second variation of the potential, corresponding to the quantization of the normal modes around that background.

The quantization problem is now more subtle: it is not simply defined by the holonomies that enclose quantized units of flux of the curvature $F=dA$. New possibilities arise that exist neither for a free particle nor for a pure interaction term (\ref{eA}) separately. The general quantum mechanical behavior for the system described by adding (\ref{k}) and (\ref{eA}) for arbitrary electromagnetic potentials is an open problem, but it is completely solvable for sufficiently simple $A$ (e.g., static, spherically symmetric, etc.).

The introduction of a kinetic term requires a new ingredient that is not necessary for the interaction term: a spacetime metric $\gamma_{ab}(z)$. This object describes some additional feature of the system, namely the metric properties of the space with coordinates $\{z^a\}$. In many cases, the metric is a prescribed non-dynamical function, while in more realistic settings like in General Relativity $\gamma_{ab}(z)$ can also be a dynamical field. This, however, could be an unwanted feature in a fundamental theory of gravitation, where the spacetime metric shouldn't be given a priori, but should be an output of the dynamical system.

The parallel between a mechanical system with $s$ degrees of freedom and the (0+1)-CS action in a $(2s+1)$-dimensional target space can be summarized in the following table:

\vspace{0.5cm} 
\begin{tabular}{|c|c|}
\hline
\textbf{Classical mechanics} & \textbf{0+1 Chern-Simons} \\ \hline
Hamilton's equations & Vanishing Lorentz force\\ 
Invariance under canonical transformations & Gauge invariance\\ 
Invariance under time reparametrizations & Invariance under Gen. Coord Transf. \\ 
Bohr-Sommerfeld quantization & Flux/holonomy quantization\\ \hline
\end{tabular}
\vspace{0.5cm}

To close this section, we summarize the features of the functional (\ref{eA}) that are common to all CS actions,

\begin{itemize}
\item \textbf{Topological origin.} The CS form is related to a topological density in $2n$ dimensions known as a characteristic class $C_{2n}$, through 
\begin{equation}  \label{CS}
dL_{2n-1}(\mathbf{A})=C_{2n}(\mathbf{A}).
\end{equation}

\item \textbf{No metric required.} Since the connection is a local 1-form, it is ready for integrating it over $\Gamma$, without any additional structure. The metric of $M^{2s+1}$ is irrelevant, which is a consequence of the topological nature of the CS forms. This continues to be the case if instead of the 1-dimensional manifold $\Gamma$, the action is defined on a higher-dimensional world volume.

\item \textbf{Gauge quasi-invariance.} The functional (\ref{eA}) involves explicitly the connection $A$ and cannot be expressed as an integral of a
gauge invariant local function. However, under a gauge transformation $A\rightarrow A+ d\Omega$, $I$ changes by the surface term 
\begin{equation}  \label{deltaI}
\delta I = e\int_{\Gamma} d\Omega = e[\Omega(z_+)-\Omega(z_-)],
\end{equation}
where $z_+$ and $z_-$ are the end points of the worldline $\Gamma$. Thus, the action is not strictly gauge invariant, but \textit{quasi} invariant. However, $I$ is a genuine gauge invariant object for the class of gauge transformations that vanish at the end points, $\Omega(z_+)= 0$, and $\Omega(z_-)=0$. But of course, it is not necessary to fix $\Omega=0$ at both ends; it is sufficient to require $z$ to be a periodic function, so that $\Omega(z_+)= \Omega(z_-)$, which is a again an indication that $\Gamma$ should closed (without boundary).

\item \textbf{Nontrivial dynamics.} Even though there is no ``kinetic term" ($\dot{z}^2 /2$) in $I$, the lagrangian defined by (\ref{eA}) gives a meaningful action principle. Indeed, varying with respect to $z$ one obtains the first order equations 
\begin{equation}  \label{Fz}
F_{ab} \dot{z}^b=0,
\end{equation}
which describe the motion of a charge under the influence of an external electromagnetic field such that the electric and magnetic forces exactly cancel each other. Varying with respect to $A$ informs us that this field can only vary as a connection, but is otherwise arbitrary. In higher dimensions, the field $\mathbf{A}$ is not an arbitrary connection without dynamics. The variation of the action with respect to it yields a set of field equations for $\mathbf{A}$ on the (2n+1)-dimensional world volume.
\end{itemize}

\section{Chern-Simons Field Theories}
A standard CS lagrangian is a $2n+1$ form constructed for a connection $A$ and its exterior derivatives in a way that cannot be written as a local
polynomial in curvatures. The simplest example of such expressions is the three-form $\mathcal{C}_3(A)=A\wedge dA $, where $A$ is a connection for an abelian group. The action that corresponds
to (\ref{eA}) now reads 
\begin{equation}  \label{CS3}
I[\mathbf{A}]= e \int_{\Gamma^{3}} A\wedge dA,
\end{equation}
where $\Gamma^{3}$ denotes a three-dimensional world volume swept by the evolution of a 2-dimensional membrane. Note that since the Lagrangian now
involves derivatives of the connection, extremizing the action under variations of $A$ yields dynamical field equations for this field. That wasn't the case in the preceding section, where $A(z)$ was a prescribed, non dynamical function (an external potential). Another novelty that occurs when the worldvolume $\Gamma$ has $2n+1$ dimensions is the possibility of including nonabelian connections. A nonabelian connection $\mathbf{A}$ is a 1-form with values in a nonabelian Lie algebra $\mathcal{G}$.

The three-dimensional Chern-Simons form for a nonabelian Lie algebra $\mathcal{G}$, is
\begin{equation}  \label{2+1CS}
 \mathcal{C}_3[A]=\langle \mathbf{A}\wedge d\mathbf{A} +\frac{2}{3} \mathbf{A}\wedge \mathbf{A}\wedge \mathbf{A} \rangle,
\end{equation}
where $\langle \cdots \rangle$ denotes the symmetrized trace\footnote{A note of caution is in order here: The symmetrized trace $\langle \cdots \rangle$ is a two-entry object, $\langle , \rangle:  \mathcal{G}\times \mathcal{G} \to \mathbf{R}$. So, the expression $\langle \mathbf{A} \wedge \mathbf{A}\wedge \mathbf{A} \rangle$ actually stands for $1/2 A^a\wedge A^b \wedge A^c \langle \mathbf{G}_a, [\mathbf{G}_b, \mathbf{G}_c] \rangle$, where $\mathbf{G}_a$ are the generators of the Lie algebra $\mathcal{G}$. Naturally, if $\mathcal{G}$ is abelian this term vanishes identically.} over $\mathcal{G}$. 

The defining property of this expression is the fact that its exterior derivative is a topological density in four dimensions. In other words, its exterior derivative is a four form whose integral is a topological invariant of the four manifold over which it is integrated, 
\begin{equation}
d\mathcal{C}_3= \langle \mathbf{F} \wedge \mathbf{F} \rangle,
\end{equation}
as can be directly checked in this case. Varying the CS action with respect to $\mathbf{A}$ produces --rather simple but non trivial-- field equations, 
\begin{equation}
\langle G_k \mathbf{F} \rangle=0,
\end{equation}
where $G_k$ are all the generators in the algebra. These equations mean that on any open patch, the connection is flat and can be locally written as a pure gauge,
\begin{equation}
\mathbf{A}(z)=g^{-1}dg,
\end{equation}
where $g(z)$ is any application from the worldvolume onto the gauge group, $g:\Gamma^{3}\rightarrow \mathbb{G}$. This means in particular, that there
are no propagating degrees of freedom in this theory, since any configuration is locally equivalent to a flat connection $\mathbf{A}=0$, and can be gauged away.  However, a locally flat connection doesn't necessarily describe a trivial situation. In fact, there could be many topologically nontrivial locally flat configurations, as seen for example in the case of AdS gravity in 2+1 dimensions, where the CS action for the AdS group, $SO(2,2)$, has locally flat classical solutions (constant Riemannian curvature and vanishing torsion), that describe interesting configurations such as black holes \cite{Banados:1992wn,Banados:1992gq}.

\subsubsection{(2n+1)-Dimensions}
	The extension of CS theories to higher dimensions is not trivial but straightforward. The generalization is achieved by looking for the $(2n-1)$-forms whose exterior derivative yields a Chern class for a given Lie-algebra valued connection in $(2n)$ dimensions. The construction is rather unambiguous and only requires identifying an appropriate $n$th rank invariant tensor $\tau_{k_1 k_2 \cdots k_{n+1}} := \langle G_{k_1} G_{k_2} \cdots G_{k_{n+1}} \rangle$, where $\langle \cdots \rangle$ is again a symmetrized trace over the Lie algebra. Then, the CS lagrangian form reads
\begin{equation}  \label{2n+1CS}
C_{2n+1}[\mathbf{A}]=\langle \mathbf{A}\wedge (d\mathbf{A})^n+ \alpha_1 \mathbf{A}^3 \wedge (d\mathbf{A})^{n-1} + \cdots + \alpha_{n} \mathbf{A}^{2n+1} \rangle,
\end{equation}
where $\{\alpha_1, \cdots \alpha_{n}\}$ are fixed rational numbers.

CS theories for dimensions $D\geq 5$ have been studied in different contexts (see e. g., \cite{Floreanini:1989gv,Nair-Schiff}), and in contrast with the case $D=3$, they are not necessarily topological and they do contain propagating degrees of freedom \cite{Banados-Garay-Henneaux}. An unexpected feature of these systems is that the symmetry generators (first class constraints) may become functionally dependent in some regions of phase space, called irregular sectors. When this happens, Dirac's canonical formalism cannot be applied, obscuring the analysis of the dynamical content of CS theories \cite{Henneaux:1990au}. There, the canonical analysis breaks down and it is no longer clear how to identify the physical observables (propagating degrees of freedom, conserved charges, etc.). These irregularities also imply that the theory is not correctly described by its linearized approximation and hence the perturbative analysis cannot be trusted \cite{Chandia:1998uf}. Fortunately, the troublesome configurations generically occur in sets of measure zero in phase space and one can always restrict the attention to open sets where the canonical analysis holds \cite{Miskovic:2003tn,Miskovic:2003ex}. Such canonical configurations fill most of the phase space and, as shown in \cite{Miskovic:2005di}, it is possible to
find generic (regular and non degenerate) sectors in CS theories, where the canonical formalism holds. It remains to be seen whether among the generic
configurations one can find states that could be regarded as vacua for perturbatively stable field theories \cite{Miskovic:2006ei}.

\subsection{Quantum mechanics}
The quantum behavior of the $2+1$ CS systems is quite remarkable, as pointed out by Witten \cite{Witten:1988hf}. CS theories are finite (exactly soluble), and provide a framework to understand the Jones polynomial of knot theory in three-dimensional spaces. CS theories are also related to conformal field theories in 1+1 dimensions, furnishing a precursory example of the AdS/CFT correspondence, conjectured a decade later in the context of string theories. Since there is a vast literature over the past twenty years on this subject, there is no point in repeating it here. For an intersting discussion on the relation between Wilson loops in quantum $2+1$ CS theory and knot theory, see \cite{Labastida}.

The assertion about the absence of local degrees of freedom applies to the classical solutions, so one could imagine that there might exist fluctuations around locally flat configurations that could possess a quantum spectrum of small oscillations around a stationary point of the action. However, this possibiliy doesn't exist either: a small fluctuation around a classical solution should satisfy 
\begin{equation}
\delta F=d(\delta A)=0,
\end{equation}
which means that any fluctuation around a classical configuration must be closed forms, $\delta A=d($something$)$. So, all directions in function space around a classical solution are gauge directions, and hence, contribute perturbatively nothing to the path integral. The argument is the same for non abelian as well as for abelian gauge connections.

The conclusion of this little excercise is that the quantum spectrum of 2+1 CS theories --if any-- must be nonperturbative. Following the reasoning of \cite{Witten:1988hf}, it can be seen that the quantum Hilbert space corresponds to nontrivial holonomies (i.e., Wilson loops) of the connection $\mathbf{A}$. What can be expected in higher dimensions?

For the reasons explained above, almost nothing is known about the full quantum behavior of higher-dimensional CS systems. However, all CS systems possess a vacuum configuration, $\mathbf{A}=0$, which is maximally symmetric, absolutely stable and isolated from the rest of the configurations. The theory has no local propagating degrees of freedom around this configuration, and therefore this point defines a topological field theory, like the entire CS theory in three spacetime dimensions. Indeed, in all dimensions, the perturbations around this configuration are pure gauge, $\delta \mathbf{A}=\mathbf{g}^{-1}(x) d \mathbf{g}(x)$.

Thus, the quantum spectrum of these theories is given by all possible locally flat connections defined by the group elements $\mathbf{g}: M^{2n+1}\rightarrow G$. This set is divided into equivalence classes and the quantization problem must be related to the possibility of classifying all inequivalent $\mathbf{g}(x)$s. For a pure gauge configuration $\mathbf{A}=\mathbf{g}^{-1}(x) d\mathbf{g}(x)$, the CS action takes the form of a WZ theory without the kinetic term, 
\begin{equation}
I[g] =\mbox{const}\times \int _{M^{2n+1}}\langle (\mathbf{g}^{-1} d \mathbf{g})^{2n+1} \rangle,
\end{equation}
which defines a field theory for $\mathbf{g}$ on the boundary of $M^{2n+1}$. This boundary theory inherits its symmetries from the bulk action, has propagating degrees of freedom and must also contain nonperturbative nonlocal states analogous to Wilson loops, but which in this case could be higher dimensional surfaces.

\subsection{Coupling CS systems to sources}
A physically meaningful theory must have observable quantities. The previous discussion indicates that a possible set of observables of a CS system is given by the holonomies --Wilson loops--, and their higher dimensional generalizations --Wilson surfaces. These are embedded surfaces that correspond to global gauge invariant quantities, classified by their topological nature. This is very likely all that could be done for CS theory in 2+1 dimensions, where there are no local degrees of freedom. But in higher dimensional CS theories, it should in principle be possible to couple the connection to local currents (sources), in order to probe the local degrees of freedom. Along this line, there is a host of charged $2k$-branes that can in principle couple to the connection.

The natural coupling is through a generalization of the formula (\ref{eA}), where instead of the $0+1$ CS form $A$, one can also consider any of the lower dimensional CS forms that could be embedded in the target space $M^D$, 
\begin{equation}  \label{eCS(A)}
I_{2k+1}= \int_{M^D} j^{2k+1}(x) \cdot CS(A)_{2k+1},
\end{equation}
where $j^{2k+1}$ is the current density with support on the worldvolume of the $2k$ brane, and $CS(A)_{2k+1}$ is the $2k+1$ CS form for the connection $A$ ( $0<k<D/2$).  The sole function of the current density is to project onto $\Gamma^{2k+1}$, and (\ref{eCS(A)}) can then be written as the integral of the CS form on the worldvolume of the brane, 
\begin{equation}
I_{2k+1} =\kappa \int_{\Gamma^{2k+1}} CS(A)_{2k+1}.  \label{eCS(A)'}
\end{equation}
One can consider a scenario where a $2n+2$-CS field theory couples to different $2k$-dimensional branes for $k=0,1, 2,...2n-k$, the simplest example being that of a zero-brane sitting in a $2+1$-CS theory \cite{Mora-Nishino}. In the case of $2+1$ gravity with a one-dimensional topological defect, the source is a point mass whose magnitude is proportional to the defect produced by an identification around the worldline. The geometry generated is that of a $2+1$ black hole \cite{Banados:1992wn,Banados:1992gq} of a certain mass and angular momentum \cite{Miskovic-Z08}. The examination of this model in higher dimensions and for a supersymmetric system  will be discussed elsewhere \cite{Edelstein-Z08}. 

\textbf{\Large Acknowledgments}\newline
This work was supported by Fondecyt grant \# 1020629. The Centro de Estudios Cient\'{\i}ficos (CECS) is funded by the Chilean Government through the
Millennium Science Initiative and the Centers of Excellence Base Financing Program of Conicyt. CECS is also supported by a group of private companies
which at present includes Antofagasta Minerals, Arauco, Empresas CMPC, Indura, Naviera Ultragas and Telef\'onica del Sur.


\end{document}